\documentstyle[11pt, titlepage]{article}

\newcommand{\beq}{\begin{equation}}
\newcommand{\beqn}{\begin{eqnarray}}
\newcommand{\eeq}{\end{equation}}
\newcommand{\eeqn}{\end{eqnarray}}
\begin{document}
\begin{titlepage}
\rightline {Preprint DART-HEP 93/06}
\rightline {astro-ph/9308043}
\rightline {Forthcoming in {\it Phys. Rev. D}}
\begin{center}
\bigskip \bigskip \bigskip \bigskip \bigskip
\Large\bf Constraints in the context of\\
Induced-gravity Inflation\\
\bigskip\bigskip\bigskip
\normalsize\rm
David I. Kaiser \\
\bigskip
\it Department of Physics\\
Harvard University\\
Cambridge MA 02138 \\ \rm
e-mail:  dkaiser@husc.harvard.edu\\
\bigskip\bigskip\bigskip\bigskip
17 February 1994\\
\bigskip
[Revised Draft] \\
\bigskip\bigskip\bigskip\bigskip\bigskip\bigskip
\bf Abstract \rm
\end{center}
\narrower Constraints on the required flatness of the
scalar potential $V(\phi)$ for a cousin-model to extended inflation
are studied.  It is shown that, unlike earlier results,
Induced-gravity Inflation can lead to successful inflation with
a very simple lagrangian and $\lambda \sim 10^{-6}$, rather than
$10^{-15}$ as previously reported.  A second order phase transition
further enables this model to escape the \rm\lq big bubble' problem of
extended inflation, while retaining the latter's motivations based on
the low-energy effective lagrangians of supergravity, superstring, and
Kaluza-Klein theories.\bigskip
\end{titlepage}
\newpage

\baselineskip 28pt
\section{Introduction}
\indent  Since Guth's first paper on inflation a decade ago~\cite{1},
there has been an explosion of effort dedicated to developing a natural
theory of inflationary cosmology.  These efforts have established
inflation as a ubiquitous program of research among cosmologists and
particle physicists, but have not yet produced any particular model
which satisfies all of the practitioners.  As Kolb and Turner recently
remarked, inflation is by now \rm\lq\lq a paradigm in search of a
model."~\cite{2}  Much of the trouble concerns the so-called \rm\lq\lq
fine-tuning" required to force the various models into agreement with
measurements of the small anisotropy of the cosmic microwave background
radiation.\\
\indent  Adams, Freese, and Guth~\cite{3} developed a quantitative
means of measuring the amount of \rm\lq\lq fine-tuning" necessary
to make inflationary models agree with observations.  They define a
dimensionless parameter $\Lambda$ as the ratio of the change in the
scalar field's potential to the change in the scalar
field:
\beq
\Lambda \equiv \frac{\Delta V(\phi)}{\left(\Delta \phi\right)^4}.
\eeq
In this way, $\Lambda$ functions as a measure of the flatness of a given
potential.  (Ref.~\cite{3} denotes this ratio by $\lambda$, rather
than $\Lambda$; the upper-case letter is used here to avoid
confusion with the closely-related quartic self-coupling parameter.)
In~\cite{3}, they evaluate this ratio for generic
inflation scenarios (without a curvature-coupled scalar field)
and for the original version of extended inflation~\cite{4}.  They show
that extended inflation requires a fine-tuning $eight$ orders of
magnitude more stringent than the general inflation schemes:
whereas they find $\Lambda \leq 10^{-6} - 10^{-8}$ for the inflaton
potential of new inflation,
they calculate that a potential for the Brans-Dicke-like scalar field in
extended inflation would require $\Lambda \leq 10^{-15}$.  Although
the authors of~\cite{3} are quick to point out that their method of
calculating $\Lambda$ is highly model-dependent, others have taken this
result to indicate that adding a potential $V(\phi)$ for the
curvature-coupled scalar in $any$ type of extended inflation model
necessarily entails this sort of extreme fine-tuning (see, $e.g.$,
{}~\cite{5}).\\
\indent  This paper examines the constraints on $V(\phi)$ for a
cousin-model to the original extended inflation.  Building on Zee's
early ideas~\cite{6} about uniting spontaneous symmetry breaking with the
Brans-Dicke reformulation of general relativity, Accetta,
Zoller, and Turner~\cite{7} developed a model of \rm\lq\lq
Induced-gravity Inflation."  In this model, a single scalar boson does
all the work of inflation:  it couples to the scalar curvature $R$ $and$
drives inflation with $V(\phi)$ (unlike ordinary extended inflation,
which requires one field to couple to $R$ while a separate and unrelated
field drives the expansion).  This kind of one-boson model can agree
with observations with about the same degree of \rm\lq\lq fine-tuning"
as the generic
models of inflation examined in~\cite{3}:  as will be shown below,
\rm\lq\lq Induced-gravity Inflation" requires $\Lambda \leq
\cal{O}\rm(10^{-6})$ -- a far cry from the $10^{-15}$ of ref.~\cite{3}!
Furthermore, by employing a $second$ order phase transition to exit the
inflationary epoch, rather than the first order transition of~\cite{4},
the \rm\lq big bubble' or \rm\lq $\omega$ problem' which plagued
original extended inflation~\cite{9,10} may be avoided.\\
\indent  In section 2, we calculate $\Lambda$ for Induced-gravity
Inflation, and consider why earlier attempts to determine the required
flatness of the potential have led to much more constrained results.
Section 3 examines the accuracy of the slow-rollover solutions upon
which the calculation of $\Lambda$ is based.  And in section 4, we
briefly consider benefits and difficulties of placing the inflationary
epoch at such a high energy scale.\\
\section{Calculating $\Lambda$}
\indent  We begin with the lagrangian density:\footnote{The sign
conventions follow those of Misner, Thorne, and Wheeler~\cite{14}, which
are based upon the 1962 edition of Landau and Lifshitz~\cite{14}.  Thus,
$g_{tt} < 0$, the full Riemann tensor is $R^{\lambda}_{\mu\nu\kappa} =
\partial_{\nu}\Gamma^{\lambda}_{\mu\kappa} -
\partial_{\kappa}\Gamma^{\lambda}_{\mu\nu} +
\Gamma^{\lambda}_{\sigma\nu} \Gamma^{\sigma}_{\mu\kappa} -
\Gamma^{\lambda}_{\sigma\kappa} \Gamma^{\sigma}_{\mu\nu}$, and the Ricci
tensor is $R_{\mu\kappa} \equiv R^{\lambda}_{\mu\lambda\kappa}$.  The
original Brans-Dicke papers followed these sign conventions.  Note that
these definitions for the Riemann and Ricci tensors are opposite in sign
from those in Weinberg's text~\cite{14} (and thus opposite to some of
the recent literature on extended inflation).}
\beq
\cal{L}\rm = \it f(\phi)R \rm - \frac{1}{2} \it g^{\mu\nu} \partial_{\mu}
\phi \partial_{\nu}\phi - V(\phi) + \it L_M \rm ,
\eeq
where the $L_M$ term here only includes contributions from \rm\lq
ordinary' matter -- there is no separate Higgs sector as in~\cite{4}.
If we choose $f(\phi) = \phi^2/(8\omega)$, we find the coupled field
equations:
\beqn
\nonumber
H^2 + \frac{k}{a^2} &=& \frac{4\omega}{3 \phi^2} \left( \rho + V(\phi) \right)
+
\frac{2\omega}{3}\left(\frac{\dot{\phi}}{\phi}\right)^2 -
2H\left(\frac{\dot{\phi}}{\phi}\right) ,\\
\ddot{\phi} + 3H\dot{\phi} + \frac{\dot{\phi}^2}{\phi} &=&
\frac{2\omega}{3 + 2\omega} \phi^{-1} \left[ (\rho - 3p) + 4V(\phi)
- \phi V^{\prime}(\phi)\right].
\eeqn
In eq. (3), $a(t)$ is the cosmic scale factor of the Robertson-Walker
metric, and is related to the Hubble parameter by $H \equiv \dot{a} /
a$.  From eq. (2), one can see that $f(\phi) \rightarrow (16\pi
G_{eff})^{-1}$, which leads to $4\omega \phi^{-2} = 8\pi G_{eff}$.
  Note that we have parametrized the lagrangian slightly differently
from the 1985
\rm\lq\lq Induced-gravity" paper:  our Brans-Dicke parameter $\omega$ is
inversely proportional to their coupling strength $\varepsilon$ ($\omega
= (4\varepsilon)^{-1}$).  That paper also makes the minor approximation
that $(2\omega) / (3 + 2\omega) \rightarrow 1$, whereas we have kept
this term explicit.  During the inflationary epoch, the $k$ term becomes
negligible;
similarly, since $\rho$ and $p$ now only include contributions from
ordinary matter, they too may be neglected.  If we now impose the
appropriate slow-rollover approximations~\cite{7,8}:
\beqn
\nonumber \left|\frac{\dot{\phi}}{\phi}\right| &\ll& H,\\
\nonumber \left|\ddot{\phi}\right| &\ll& 3H\dot{\phi},\\
\left|\dot{\phi}^2\right| &\ll& V(\phi),
\eeqn
the field equations enter a more tractable form:
\beqn
\nonumber H^2 &\simeq& \frac{4\omega}{3} \frac{V(\phi)}{\phi^2},\\
3H\dot{\phi} &\simeq& \frac{2\omega}{3 + 2\omega}\frac{\left[ 4 V(\phi)
- \phi V^{\prime}(\phi)\right]}{\phi}.
\eeqn
In order to calculate the required flatness ($\Lambda$) of the potential,
we must now assume a particular form for $V(\phi)$.  To achieve a
second order phase transition with a minimum of
fine-tuning, we may choose the simplest form for the potential at
tree-level:
\beq
V(\phi) = \frac{\lambda}{4} \left(\phi^2 (t) - v^2\right)^2,
\eeq
where $\lambda$ is the quartic self-coupling, and $v$ is the vacuum
expectation value for $\phi$.  In addition to being a qualitatively
simple form for the desired potential, this is also the optimal
form found in~\cite{3} for the case of generic new inflation.  It also
matches the potential of~\cite{7}, with the minor difference that it is
parametrized with $\lambda / 4$ rather than $\lambda / 8$.\\
\indent  Using this expression for $V(\phi)$, combined with the field
equations (5), we may solve for $\phi(t)$ and $a(t)$:
\beqn
\nonumber \phi(t) &=& \phi_{\circ} + \left(\frac{\lambda \omega}{3
\gamma^2}\right)^{1/2} v^2 t,\\
\frac{a(t)}{a_B} &=& \left(\frac{\phi(t)}{\phi_{\circ}}\right)^{\gamma}
exp\left[ \frac{\gamma}{2 v^2}\left(\phi^2_{\circ} -
\phi^2(t)\right)\right].
\eeqn
The factor $\gamma \equiv (3 + 2\omega) / 2$ is slightly different
from the exponent found in~\cite{7} (because of their earlier
approximation that $(2\omega) / (3 + 2\omega) \rightarrow 1$).  The
quantities $\phi_{\circ}$ and $a_B$ are values at the beginning of the
inflationary epoch.  Note that at early times, when $\phi(t) \sim
\phi_{\circ}$,
these equations yield the familiar power-law solution, with $a(t) \propto
t^{\gamma}$.  As $t$ increases, the rate of expansion slows due to the
$exp\> (\phi^2_{\circ} - \phi^2)$ term.  With these
analytic expressions for $\phi(t)$ and $a(t)$, we may now calculate
$\Lambda$.\\
\indent  Following~\cite{3}, we begin with the two basic constraints on
the scalar potential:  it must provide for sufficient inflation to solve
the flatness, horizon, and monopole density problems, and it must allow
for the proper amplitude of density perturbations to act as seeds for
the evolution of large scale structure.  The first of these requirements
takes the form:
\beq
\frac{1}{H_N\>a_N} \leq \frac{1}{H_B\>a_B},
\eeq
where the subscript $N$ refers to present values (\rm\lq Now'), and $B$
refers to values at the beginning of the inflationary epoch.  Using
the standard cosmological model's assumption of adiabatic expansion
following the end of inflation, we may write:
\beq
\frac{a_{end}}{a_N} \simeq \frac{T_N}{T_{RH}},
\eeq
where $a_{end}$ is the value of the scale factor at the end of
inflation, and $T_{RH}$ is the reheat temperature following
thermalization of the foregoing false vacuum energy density.
(Ref.~\cite{3} assumes $T_{RH} \approx M_F$, where $M_F^4$ is the false
vacuum energy density of the second boson which drives inflation.)
The Hubble parameter may then be parametrized as:
\beqn
\nonumber H_B^2 &=& \frac{8\pi}{3}\>\frac{V_f}{m_p^2},\\
H_N^2 &=& \frac{8\pi}{3}\>\frac{\beta^2\>T_N^4}{M_p^2},
\eeqn
where $V_f$ is the initial value of the false vacuum energy density;
$m_p$ is the effective value of the Planck mass at the beginning of
inflation (related to the initial value of the field $\phi$:
$\phi_{\circ} = (\omega / 2\pi)^{1/2} m_p$); and $M_p$ is the present
value of the Planck mass, $M_p \approx 1.22 \times 10^{19}$ GeV.
The quantity $\beta$ is the ratio of the energy density in matter to the
energy density in radiation today, which~\cite{3} takes to be around 81.
In the original extended inflation
framework, $V = V_f = M_F^4$ was constant, due to the energy density of the
metastable state of the second boson before it completed its first order
phase transition.  In the present model, $V(\phi)$ changes in time as
$\phi(t)$ changes; yet even in this new
context, $V_f$ is still a constant, and is simply equal to
$V(\phi_{\circ})$; its use here is thus independent of any slow-roll
approximation.\\
\indent  Using eqs. (8-10), the condition for sufficient inflation takes
the form:
\beq
\frac{a_{end}}{a_B} \geq \frac{V_f^{1/2}}{\beta\>T_N\>T_{RH}}\>
\frac{M_p}{m_p}.
\eeq
Combining with eq. (7), we get:
\beq
\nonumber \frac{a_{end}}{a_B} = \left(\frac{\phi_{end}}{\phi_{\circ}}
\right)^{\gamma} exp \left[\frac{\gamma}{2v^2} \left( \phi_{\circ}^2 -
\phi_{end}^2 \right) \right]
\geq
\frac{V_f^{1/2}}{\beta\>T_N\>T_{RH}} \> \frac{M_p}{m_p},
\eeq
where we have followed~\cite{7} in utilizing the explicit analytic
solution for $a(t)$ based on the slow-roll approximation right up to
$a_{end}$; as we shall see below, this approximation is a good one:  the
slow-rollover solutions employed above don't break down until $\phi(t)
\simeq 0.98v$.\\
\indent  Eq. (12) may be used to place a bound on the change in the
scalar field, $\Delta \phi = \left( \phi_{end} - \phi_{\circ}\right)$.
Writing $\phi_{\circ}$ and $\phi_{end}$ in
the exponent in terms of their associated masses, we find:
\beq
\left(\frac{\phi_{end}}{\phi_{\circ}}\right) \geq \left[ \frac{
V_f^{1/2}}{\beta\>T_N\>T_{RH}} \> \frac{M_p}{m_p}\right]^{1/\gamma} \>
exp \left[\frac{-\omega}{4\pi v^2}\>\left(m_p^2 - M_E^2 \right)\right],
\eeq
where we have followed~\cite{3} in writing $M_E$ for the value of
the Planck mass at the end of inflation.  In the simplest case, $\phi$
would not evolve any more after the end of inflation, so $M_E$
would equal $M_P$; yet for the time being the more general value $M_E$
will be used.  Since $\phi_{\circ} = (\omega / 2\pi)^{1/2} m_p$ and $v
\simeq \phi_{end} = (\omega /2\pi)^{1/2} M_E$, eq. (13) may
be rewritten:
\beq
\Delta \phi \geq \sqrt{\frac{\omega}{2\pi}} \>
m_p \left( \left[
\frac{V_f^{1/2}}{\beta \> T_N \> T_{RH}} \>
\frac{M_p}{m_p}\right]^{1/ \gamma}\>
 exp \left[ \frac{1}{2}\left( 1 -
\left( \frac{m_p}{M_E} \right)^2 \right) \right] - 1 \right).
\eeq
Taking $\Delta V = V_f$ and defining $\mu \equiv (m_p / M_E)$,
 the ratio $\Lambda$ thus becomes:
\beq
\Lambda \leq \left(\frac{2\pi}{\omega}\right)^2 \> \frac{V_f}{m_p^4} \>
\left(\left[
\frac{V_f^{1/2}}{\beta\>T_N\>T_{RH}} \>
\frac{M_p}{m_p}\right]^{1/ \gamma}\> exp\left[ \frac{1}{2}\left( 1 -
\mu^2 \right)\right] - 1\right)^{-4}.
\eeq
Incorporating the constraint that $V_f$ must exceed the kinetic energy
of the $\phi$ field's de Sitter space fluctuations without exceeding the
effective Planck scale at the beginning of inflation~\cite{3}, we may
bound $V_f$ by:
\beq
V_f \leq \left(\frac{3}{8\pi}\right)^2 m_p^4,
\eeq
which may then be used to eliminate the coefficient of $V_f / m_p^4$:
\beq
\Lambda \leq \left(\frac{3}{4\omega}\right)^2 \>
\left(\left[
\frac{V_f^{1/2}}{\beta\>T_N\>T_{RH}} \>
\frac{M_p}{m_p}\right]^{1/ \gamma}\> exp\left[ \frac{1}{2}\left( 1 -
\mu^2 \right)\right] - 1\right)^{-4}.
\eeq
\indent  Thus far we have only relied upon the requirement of sufficient
inflation.  The second requirement, based on the amplitude of density
perturbations, may now be used to place a bound on the ratio $T_{RH}^2 /
V_f^{1/2}$.  This bound can then be used to remove nearly all mass-scale
dependence from $\Lambda$.  The post-COBE parametrization of the density
perturbation constraint may be written as~\cite{11}:
\beq
\left(\frac{H^2}{\dot{\phi}}\right)_{\vert hor} \leq 5\pi \delta_H,
\eeq
where the quantity $H^2 / \dot{\phi}$ is to be evaluated at the time of
last horizon-crossing of the
perturbations (during the inflationary epoch), and
 $\delta_H \simeq 1.7 \times 10^{-5}$.  (Ref.~\cite{3}
used $5 \times 10^{-4}$, rather than $5\pi \delta_H$.)  As noted in
{}~\cite{12}, the quantities pertaining to the time of last
horizon-crossing in eq. (18) should be evaluated in the Einstein
frame; when
compared with the \rm\lq\lq naive" calculations in the Jordan frame, a
correction factor $F(\omega)$ must be introduced.  Yet this factor
$F(\omega)$ decreases monotonically with $\omega$, and converges to
unity rather quickly:  $F(\omega = 25) = 1.052$, and $F(\omega = 500) =
1.002$.  For the particular values of $\omega$ with which
we shall be concerned below, $F(\omega)$ would thus be negligible, and
so we may continue to calculate $H^2 / \dot{\phi}$ in the Jordan frame.
(This is also the approach
adopted in~\cite{5}.)  Since we want to relate eq. (18) to a bound
on $V_f$, we should first rewrite $\dot{\phi}(t_*)$ in terms of $V_f$,
where $t_*$ is the time of horizon-crossing.  From eq. (7) we have:
\beq
\dot{\phi}(t) = \left(\frac{\lambda \omega}{3 \gamma^2}\right)^{1/2} \>
\frac{\omega}{2\pi} \> M_E^2,
\eeq
and from eq. (6) we have:
\beq
V_f = V(\phi_{\circ}) = \frac{\lambda \omega^2}{16\pi^2}\> M_E^4 \>
\left(1 -  \mu^2 \right)^2 .
\eeq
Using these expressions, we may then rewrite (the
time-independent) $\dot{\phi}$ in terms of the constant $V_f$:
\beq
\dot{\phi} = \left(\frac{4\omega}{3 \gamma^2}\right)^{1/2}
V_f^{1/2} \> \left| \left(1 - \mu^2 \right)^{-1}\right|.
\eeq
The absolute value for the $\left( 1 - \mu^2\right)$ term comes from
taking the positive square root of $M_E^4$.  Since $m_p < M_E$, the pole
at $\mu^2 = 1$ is excluded.\\
\indent  Now we need to calculate the Hubble parameter at the
time $t_*$. Following~\cite{3}, we may write:
\beq
\frac{1}{H(t_*)\> a(t_*)} = \frac{1}{H_N\> a_N},
\eeq
which, after employing eqs. (9, 10, 21), leads to:
\beq
\frac{H^2(t_*)}{\dot{\phi}(t_*)} \simeq
 \frac{8\pi}{3} \>
\frac{\beta^2 \> T_N^2 \> T_{RH}^2}{M_p^2 \> V_f^{1/2}} \>
\sqrt{\frac{3\gamma^2}{4\omega}} \> \left(\frac{a_{end}}{a(t_*)}
 \right)^2\> \left(1 - \mu^2 \right) \leq 5\pi \delta_H.
\eeq
Rewriting this as a bound on $(T_{RH}^2 / V_f^{1/2})$, we get:
\beq
\left( \frac{T_{RH}^2}{V_f^{1/2}}\right) \leq \frac{5 \delta_H\>
M_p^2}{4 \beta^2\> T_N^2}\> \sqrt{\frac{3\omega}{\gamma^2}} \> \left(
\frac{a_{end}}{a(t_*)}\right)^{-2}\> \left|\left(1 - \mu^2 \right)^{-1}
\right|.
\eeq
This may be substituted into eq. (17) for $\Lambda$:
\beq
\Lambda \leq \left(\frac{3}{4\omega}\right)^2 \> \left(\left[
\frac{4 \> \beta \> T_N}{5\> \delta_H \> M_p} \> \frac{T_{RH}}{m_p}
\> \sqrt{\frac{\gamma^2}{3 \omega}} \>
\left(\frac{a_{end}}{a(t_*)}\right)^2 \> \left( 1 - \mu^2 \right)
 \right]^{1/\gamma} \> exp
\left[\frac{1}{2}\left(1 - \mu^2 \right)
\right] - 1 \right)^{-4}.
\eeq
We have taken the equality in eq. (24) as a worst case for $\Lambda$; if
$\left(T_{RH}^2 / V_f^{1/2}\right)$ were much less than the right-hand
side of eq. (24), the bound on $\Lambda$ would increase (and the
resultant need for \rm\lq\lq fine-tuning" would therefore decrease).
The remaining factor of $T_{RH} / m_p$ may now be removed by combining
eqs. (16, 24).  Following these substitutions, the expression for
$\Lambda$ becomes:
\beq
\Lambda \leq \left(\frac{3}{4\omega}\right)^2 \> \left( \left[
\sqrt{\frac{1}{10\pi \delta_H}} \> \left(\frac{3
\gamma^2}{\omega}\right)^{1/4} \> \left(\frac{a_{end}}{a(t_*)}\right)\>
\left(1 - \mu^2 \right)^{1/2}
\right]^{1/ \gamma} \> exp
\left[\frac{1}{2}\left(1 - \mu^2 \right) \right] - 1 \right)^{-4}.
\eeq
The time of last horizon-crossing can be calculated from eq. (22), and
from $t_*$ one could then find the ratio $a_{end} / a(t_*)$.  Yet the
dependence of $\Lambda$ on $a_{end} / a(t_*)$ is very weak (being
suppressed by the exponent $\gamma^{-1}$), so some simplifying
approximations may be made.  In most models of inflation, $t_*$ is
around 60 e-folds before the end of inflation (although some recent
models have $t_* \sim 50$ e-folds before the end of inflation,
$e.g.$~\cite{13}); this means that
the ratio $a_{end} / a(t_*)$ is simply $e^{60}$ (or, perhaps, $e^{50}$).
For the calculation of $\Lambda$, we will assume $a_{end} / a(t_*) \sim
e^{60}$ in this model.
As Figure 1 shows, $\Lambda$ increases monotonically with increasing
$\mu$ for a given value of $\omega$; a $lowest$ bound on $\Lambda$ thus
comes from taking the limit $\mu \rightarrow 0$ ($i.e.$, $m_p \ll M_E$).
 When this
is done, $\Lambda$ becomes a function of $\omega$ alone.  Figure 2 shows
a plot of $\Lambda$ versus $\omega$ in the limit $\mu \rightarrow 0$:
  the maximum value of $\Lambda$
(corresponding to the least amount of \rm\lq\lq fine-tuning" required)
is $5.4 \times 10^{-6}$, for the value $\omega_{cr} = 240$.  We can
check the dependence of $\Lambda$ on $a_{end} / a(t_*)$ by defining a
parameter $\alpha$ as $a_{end} / a(t_*) = e^{\alpha}$.  Figure 3 shows a
plot of $\Lambda$ versus $\alpha$ in the limit $\mu \rightarrow 0$ for a
particular value of $\omega$ ($\omega = 500$).
The dependence on $\alpha$ is indeed
weak:  $\Lambda$ evaluated at $(\alpha = 50, \omega = 500)$ gives $4.5
\times 10^{-6}$, whereas $\Lambda$ evaluated at $(\alpha = 60, \omega =
500)$ gives $3.8 \times 10^{-6}$.  Similarly, for $\alpha = 50$ rather
than 60, $\Lambda(\omega_{cr})$ = $7.3 \times 10^{-6}$, instead of $5.4
\times 10^{-6}$.  \\
\indent  In the original model of extended inflation, $\omega$ was
constrained to be less than 25 in order to avoid observable
inhomogeneities coming from the large range in bubble sizes, even though
present tests of Brans-Dicke gravitation versus general relativity limit
$\omega$ to the range $\omega \geq 500$ (hence the \rm\lq big bubble' or
\rm\lq $\omega$ problem' of old extended inflation).  Yet in the present
model, the second order phase transition lifts this constraint on
$\omega$; $\omega$ can now be as large as necessary to meet the
experimental limits. As one can see in Figure 2, $\Lambda(\omega)$ falls
off slowly from its maximum with increasing $\omega$.   It is
 interesting to note that $\Lambda_{max}$ occurs within a factor
of 2 of the value $\omega = 500$. Because of its agreement with present day
observations, and its proximity to $\omega_{cr}$, $\omega = 500$ appears
to be a good candidate for the Brans-Dicke parameter.\\
\indent  Eq. (1) may be used to relate $\Lambda$ to the quartic
self-coupling constant $\lambda$.  For the form of $V(\phi)$ considered
here, we find:
\beq
\Lambda = \frac{\Delta V}{(\Delta \phi)^4} = \frac{\lambda}{4} \frac{
(\phi_{\circ}^2 - v^2 )^2}{ (\phi_{\circ} - v)^4},
\eeq
or, if we keep terms only up to $\cal{O}\rm(\mu)$,
\beq
\Lambda \simeq \frac{\lambda}{4} \frac{\left[ 1 + \cal{O}\rm \left(
\mu^2 \right) \right]}{\left[ 1 -
4 \mu + \cal{O}\rm \left( \mu^2 \right) \right]}.
\eeq
Thus, $\lambda$ is of the same order of magnitude as $\Lambda$, which,
for the model under study, means $\lambda \sim \cal{O} \rm (10^{-6})$.
This value of $\lambda$ is much larger than the results in~\cite{3} for
the original model of extended inflation, indicating far less of a need
for \rm\lq\lq fine-tuning".\\
\indent  We should pause here to consider why this relatively large
value for $\Lambda$ has not been noted before.  The most important
reason is because the calculation of $\Lambda$ is highly
model-dependent.  Both papers of~\cite{3}, for example, assumed
that a separate Higgs sector would drive inflation; this meant that
their $V_f^{1/4} = M_F$ was constrained to lie at the GUT scale, with
such
ratios as $(M_F / v) \sim 10^{-5}$.  Furthermore, by insisting
upon a first order phase transition in the Higgs sector, $\omega$ was
constrained to $\omega \leq 25$.  It is interesting to note that an
attempt in 1989 to unite the original Induced-gravity Inflation model
with extended inflation~\cite{16} similarly relied upon a separate Higgs
sector to drive inflation until it underwent a first order phase
transition.\\
\indent  The authors of the 1985 paper introducing Induced-gravity
Inflation~\cite{7} studied constraints on the quartic self-coupling $\lambda$,
based also on the twin requirements of sufficient inflation and a proper
amplitude for density perturbations.  Yet their result indicated that
$\lambda \leq 10^{-14}$ for $\omega \sim 500$.  Several factors help to
explain this low result.  First, their pre-COBE parametrization of the
amplitude of density perturbations leads to an increase of an order of
magnitude for $\lambda$ when compared with present, post-COBE values.
Most important, however, is their approach to bounding $\lambda$:  they
solved for $\lambda$ in terms of the ratio $(v / \phi(t_*))$, where
$\phi(t_*)$ is the value of the field at the time of last
horizon-crossing.  Their result for $\lambda$, which in their analysis
is proportional to $sinh^{-4}[\ln (v / \phi(t_*))]$, is thus $very$
sensitive to the value of $\phi(t_*)$.  Because this value cannot be
solved exactly (even in the slow-roll approximation), two extreme
limiting regimes were studied:  $\phi(t_*) \approx v$ versus $\phi(t_*)
\ll v$.  Yet small differences in the approximation of $\phi(t_*)$ lead
to order-of-magnitude differences in their estimation of $\lambda$:  a
difference of 0.01 in the assumed value of $\phi(t_*)$ leads to a
difference in $\lambda$ of two orders of magnitude.  The method of
calculating bounds on $\Lambda$ employed in this paper avoids expanding
in terms of the unknown ratio $(v / \phi(t_*))$; the only mass ratios
involved here are of order $(\phi_{\circ} / v)^2$, and their inclusion
$raises$ the bound on $\Lambda$.  The information
regarding $\phi(t_*)$ is now contained in the ratio $(a_{end} /
a(t_*))$,
and we saw above that changing this ratio from $e^{60}$ to $e^{50}$
leads to a change in $\Lambda$ by a factor of only $\sim 1.18$ (see
Figure 3).  This
appears to be the major reason for the large split in values of
$\lambda$ between this paper and the 1985 analysis.\\
\section{Accuracy of Slow-Rollover Approximate Solutions}
\indent  We may now check the accuracy of our slow-rollover approximate
solutions by following Steinhardt's and Turner's \rm\lq\lq prescription" for
successful slow-rollover~\cite{15}.  The analysis is easiest by rewriting
eq. (2) in terms of a Brans-Dicke field $\Phi$, where $\Phi \equiv f(\phi)
= \phi^2/(8\omega)$.  The \rm\lq\lq prescription" of~\cite{15}
concerns finding conditions for when the $\ddot{\Phi}$-term may be
neglected.  In the present model, when $\ddot{\Phi}$ is negligible, the
$\Phi$ equation becomes:
\beq
\dot{\Phi} = \frac{1}{3H} \frac{1}{(3 + 2\omega)} \left[ 2 V (\Phi) -
\Phi V^{\prime} (\Phi) \right],
\eeq
where the prime now indicates differentiation with respect to $\Phi$.
Using this expression for $\dot{\Phi}$, we may calculate $\ddot{\Phi}$,
and then write the ratio $\ddot{\Phi} / (3 H \dot{\Phi})$, which
becomes:
\beq
\frac{\ddot{\Phi}}{3H\dot{\Phi}} = \frac{1}{9H^2} \frac{1}{(3 +
2\omega)} \left[V^{\prime} - \Phi V^{\prime\prime}\right] -
\frac{1}{9H^3} \frac{1}{(3 + 2\omega)} \left(\frac{\partial H}{\partial
\Phi}\right) \left[2V - \Phi V^{\prime}\right].
\eeq
{}From eq. (30), it is consistent to neglect the $\ddot{\Phi}$-term when:
\beqn
\nonumber \left| V^{\prime} - \Phi V^{\prime\prime} \right| &\ll& (3 +
2\omega) \left( 9H^2 \right),\\
\left| \left(\frac{\partial H}{\partial \Phi}\right) \> (2V - \Phi
V^{\prime}) \right| &\ll& (3 + 2\omega) \> \left( 9H^3 \right).
\eeqn
These conditions may be used to solve for when the slow-rollover
approximation breaks down; that is, solved for values of $\Phi$ for which
the left-hand side of each inequality roughly $equals$ the right-hand
side (rather than being much less than it).  Since $\Phi = \phi^2 /
 (8\omega)$, the potential for our particular model may be written
$V(\Phi) = \lambda / 4 \> (8\omega \Phi - v^2)^2$, which leads to:
\beq
\left| V^{\prime} - \Phi V^{\prime\prime} \right| \rightarrow
4\lambda\omega v^2.
\eeq
After re-expressing $H$ in terms of $\Phi$ (see eq. (5)), we find
the value of the field ($\Phi_{bd}$) for which the consistency
of the slow-roll approximation breaks down to be:
\beqn
\nonumber \Phi_{bd} &=& \frac{v^2}{8\omega} \left[ 1 +
\frac{1 - \sqrt{1 + 6\gamma}}{3\gamma} \right] \rightarrow \\
\phi_{bd} &=& v \> \left[ 1 + \frac{1 - \sqrt{1 + 6\gamma}}{3\gamma}
 \right]^{1/2}.
\eeqn
If $\omega = 500$, $\phi_{bd} = 0.98 v$. It is interesting to compare
this with the result for $\phi_{bd}$ based on the second condition for
slow-rollover.  This condition leads to the assignment:
\beq
\phi_{bd} = v \> \left[ 1 + \frac{\sqrt{72}}{288} \frac{1}{\gamma}
\right]^{1/2}.
\eeq
In other words, the second condition doesn't break down until $\phi >
v$!  Thus, the result based on the first condition will be used.\\
\indent  We may calculate a maximum reheating temperature for the model
by finding the value of $V(\phi)$ at the point where the field begins
its damped oscillations around the true minimum of the potential.
Taking this point to be $\phi_{bd}$ leads to:
\beq
V(\phi_{bd}) = \frac{\lambda v^4}{18 \gamma^2} \left( 3\gamma + 1 -
\sqrt{1 + 6\gamma} \right).
\eeq
If $\omega = 500$, $V(\phi_{bd}) = (3.2 \times 10^{-4}) \lambda v^4$, so
$T_{RH,\> max} = V^{1/4}(\phi_{bd}) = (0.13) \lambda^{1/4} v$.
Furthermore, if $\lambda \sim 10^{-6}$, then $T_{RH,\> max} = (4.1
\times 10^{-3})\> v$.  Assuming the simplest case, that $\phi$ does not
evolve after the end of inflation, then $v = \sqrt{\omega / 2\pi} M_P$,
which (for $\omega = 500$) leads to $T_{RH,\> max} \simeq 4.5 \times
10^{17}$ GeV.  We will consider possible interpretations of inflation at
this energy scale below.\\
\section{Conclusions}
\indent  Induced-gravity Inflation, which combines
properties from the \rm\lq\lq new inflation" schemes of 1982~\cite{17}
(such as a slowly rolling field leading to a second order phase
transition)
with characteristics from the original version of extended
inflation~\cite{4} (including a non-minimal $\phi \> R$ coupling),
 can lead to successful inflation with potentially
acceptable limits on \rm\lq\lq fine-tuning".  The lagrangian of eq. (2)
requires only a qualitatively simple
scalar potential associated with a single curvature-coupled scalar
field; there
is no need for adding special phenomenologically-inspired \rm\lq extra'
terms by hand to $\cal{L} \rm$, as in~\cite{5,8}.
Induced-gravity Inflation can also get all of the \rm\lq work' of
inflation done with only one boson, thereby helping to slow the
proliferation of \rm\lq\lq specialty" bosons, each of which is invented
to complete specific and unrelated tasks in the early universe.
 In addition to this simplicity, the model
retains many of the motivations for extended
inflation, based on the appearance of Brans-Dicke-like couplings in the
low-energy effective theories for various Kaluza-Klein, superstring, and
supergravity theories (see ~\cite{18}).\\
\indent  The calculation of $\Lambda$ in this paper depends only on the
amplitude of density perturbations, $\delta_H$.  Yet the character of
the spectrum of perturbations may also help to rule out various
inflationary schemes~\cite{10,13,21}.  Ordinary extended inflation, for
example, predicts a rather steep tilt away
from a scale-invariant (Harrison-Zel'dovich) spectrum of density
perturbations, which appears to contradict COBE data.  Determination of
the tensor mode contributions versus scalar modes in the density
perturbation spectrum of Induced-gravity Inflation is made more
difficult because of the deviation of eq. (7) from a simple power-law
solution, and is the subject of further study.  At early times at least,
when the evolution of $a(t)$ is roughly proportional to $t^{\gamma}$,
the present model would yield a tiny tilt away from scale-invariance:
when $a(t) \propto t^{\gamma}$, the spectral
index goes as $n = 1 - 2 / (\gamma - 1)$~\cite{10}, which in this case
(with $\omega \simeq 500$) would give $n = 0.996$.  For more on the
possibility of \rm\lq\lq observing" the inflaton potential based on the
contributions from tensor mode perturbations, see~\cite{13,21}. \\
\indent  One point of concern for Induced-gravity Inflation is the scale
at which it operates:  unlike most other inflationary schemes, which
study phase transitions associated with the breaking of a GUT symmetry
(at an energy of around $10^{14}$ to $10^{16}$ GeV),
Induced-gravity Inflation is associated with the Planck scale.  This
could lead to conflict with the value of $V(\phi)$ at the time of
last horizon-crossing.  Refs.~\cite{11,13} show that present COBE data
appear to limit $V^{1/4}(t_*) \sim (3 - 4) \times 10^{16}$ GeV, which, for
the present model (with
$\omega \simeq 500$ and $\lambda \simeq 10^{-6}$),
would require $\phi(t_*)$ to be $very$ close to $v$.  Yet, as pointed
out in~\cite{13}, uncertainties in the data lead to an entire order of
magnitude range in the value for $V^{1/4}(t_*)$, so the present model
cannot be ruled out by these COBE results.  For more on constraints on
the energy scale of inflation, see~\cite{19}.\\
\indent  A theoretical difficulty for Induced-gravity Inflation stemming
from its high energy scale is how to combine it with a \rm\lq\lq
realistic" particle physics sector.  (Recent work with extended
technicolor as a means of achieving Planck-scale unification of gauge
couplings~\cite{22} might offer a means of connecting an Induced-gravity
 Inflation model with realizable particle physics models.)  Yet what it
might lose on the particle
 side, it gains on the gravitational side:  it should be
much easier to relate the present model to a specific higher-energy
gravitational theory.  Or the model might be useful as part of
a \rm\lq\lq
Double-Inflation" scheme, in which the Induced-gravity phase transition
(which, as we have seen above, could solve the flatness and horizon
problems rather easily, and lead to an acceptable amplitude of density
perturbations)
is followed by a related GUT transition at a lower energy (which would
then only need to solve the monopole density problem, so the
requirements for this second epoch of inflation would be greatly
relaxed). (For earlier
attempts to use the original model of extended inflation in a
double-inflation scenario, see~\cite{20}.)  Although these details have
yet to be worked out, the prospect of a well-motivated inflationary
scenario which requires $\Lambda \sim 10^{-6}$ rather than $\sim
10^{-15}$ remains an encouraging result.\\
\section{Acknowledgments}
I would like to thank M. Gleiser and J. Harris for many helpful
discussions during the course of this work.\\
%


%
\newpage
\centerline{\large\bf{Figure Captions}}
\normalsize\rm
\bigskip\bigskip
\baselineskip 20pt

1.  Plot of $\Lambda$ as a function of $\mu \equiv
(m_p / M_E)$, based
on eq. (26) with $\omega = 500$.  The vertical scale is in units of
$10^{-5}$, and the assumption that $a_{end} / a(t_*) \simeq e^{60}$
has been used.  The intercept at $\mu = 0$ is $\Lambda = 3.8 \times
10^{-6}$.\bigskip\bigskip\\
 2.  Plot of $\Lambda(\omega)$ versus $\omega$,
based on eq. (26) with
$\mu \rightarrow 0$. The
vertical scale is in units of $10^{-6}$, and the
assumption that $a_{end} / a(t_*) \simeq e^{60}$ has been used.
Note that $\Lambda$ reaches its maximum value of $5.4 \times
10^{-6}$ at $\omega_{cr} = 240$; the value at $\omega = 500$ is
$\Lambda(500) = 3.8 \times 10^{-6}$.\bigskip\bigskip\\
3.  Plot of $\Lambda$ versus $\alpha$, where $a_{end} / a(t_*) =
e^{\alpha}$, based on eq. (26) with $\mu \rightarrow 0$ and $\omega =
500$.  The vertical scale is in units of $10^{-6}$.
\end{document}